%
%
\def\today{\ifcase\month\or January\or February\or March\or April\or May\or
June\or July\or August\or September\or October\or November\or December\fi
\space\number\day, \number\year}
%
%
\newcount\notenumber

\def\note{\global\advance\notenumber by 1 \footnote{$^{\the\notenumber}$}}
%
%
\newif\ifsectionnumbering
\newcount\eqnumber
\def\cleareqnumber{\eqnumber=0}
\def\numbereq{\global\advance\eqnumber by 1
\ifsectionnumbering\eqno(\the\secnumber.\the\eqnumber)\else\eqno
(\the\eqnumber)\fi}
\def\eqalinno{{\global\advance\eqnumber by 1}
\ifsectionnumbering(\the\secnumber.\the\eqnumber)\else(\the\eqnumber)\fi}
\def\name#1{\ifsectionnumbering\xdef#1{\the\secnumber.\the\eqnumber}
\else\xdef#1{\the\eqnumber}\fi}
\def\nosectionnumbering{\sectionnumberingfalse}
\sectionnumberingtrue
%
%
\newcount\refnumber

\immediate\openout1=refs.tex
\immediate\write1{\noexpand\frenchspacing}
\immediate\write1{\parskip=0pt}
\def\ref#1#2{\global\advance\refnumber by 1%
[\the\refnumber]\xdef#1{\the\refnumber}%
\immediate\write1{\noexpand\item{[#1]}#2}}
\def\tie{\noexpand~}

%
%
\font\twelvebf=cmbx10 scaled \magstep1
\newcount\secnumber

\def\newsection#1.{\ifsectionnumbering\cleareqnumber\else\fi%
	\global\advance\secnumber by 1%
	\bigbreak\bigskip\par%
	\line{\twelvebf \the\secnumber. #1.\hfil}\nobreak\medskip\par}
%
%
%
\def \sqr#1#2{{\vcenter{\vbox{\hrule height.#2pt
	\hbox{\vrule width.#2pt height#1pt \kern#1pt
		\vrule width.#2pt}
		\hrule height.#2pt}}}}

%
%
%
\newdimen\fullhsize
\def\fiddle{\fullhsize=6.5truein \hsize=3.2truein}
\def\fullline{\hbox to\fullhsize}
\def\mkhdline{\vbox to 0pt{\vskip-22.5pt
	\fullline{\vbox to8.5pt{}\the\headline}\vss}\nointerlineskip}
\def\mkftline{\baselineskip=24pt\fullline{\the\footline}}
\let\lr=L \newbox\leftcolumn
\def\twocolumns{\fiddle
	\output={\if L\lr \global\setbox\leftcolumn=\columnbox
		\global\let\lr=R \else \doubleformat \global\let\lr=L\fi
		\ifnum\outputpenalty>-20000 \else\dosupereject\fi}}
\def\doubleformat{\shipout\vbox{\mkhdline
		\fullline{\box\leftcolumn\hfil\columnbox}
		\mkftline} \advancepageno}
\def\columnbox{\leftline{\pagebody}}
\nosectionnumbering
\magnification=1200
\def\pr#1 {Phys. Rev. {\bf D#1\tie }}
\def\pe#1 {Phys. Rev. {\bf #1\tie}}
\def\pre#1 {Phys. Rep. {\bf #1\tie}}
\def\pl#1 {Phys. Lett. {\bf #1B\tie }}
\def\prl#1 {Phys. Rev. Lett. {\bf #1\tie }}
\def\np#1 {Nucl. Phys. {\bf B#1\tie }}
\def\ap#1 {Ann. Phys. (NY) {\bf #1\tie }}
\def\cmp#1 {Commun. Math. Phys. {\bf #1\tie }}
\def\imp#1 {Int. Jour. Mod. Phys. {\bf A#1\tie }}
\def\mpl#1 {Mod. Phys. Lett. {\bf A#1\tie}}
\def\tie{\noexpand~}

\def\ints{\int d\sigma\,}

\def\dx{\partial X}
\def\dbx{{\overline\partial}X}

\parskip=15pt plus 4pt minus 3pt
\headline{\ifnum \pageno>1\it\hfil O(d, d, Z) Transformations
as automorphisms
	$\ldots$\else \hfil\fi}
\font\title=cmbx10 scaled\magstep1
\font\tit=cmti10 scaled\magstep1
\footline{\ifnum \pageno>1 \hfil \folio \hfil \else
\hfil\fi}
\raggedbottom
\rightline{\vbox{\hbox{RU96-04-B}}}
\vfill
\centerline{\title O(d, d, Z) TRANSFORMATIONS AS}
\vskip 20pt
\centerline{\title AUTOMORPHISMS OF THE OPERATOR ALGEBRA}
\vfill
{\centerline{\title Ioannis Giannakis${}^{a}$ \footnote{$^{\dag}$}
{\rm e-mail: giannak@theory.rockefeller.edu}}
}
\medskip
\centerline{$^{(a)}${\tit Physics Department, Rockefeller
University}}
\centerline{\tit 1230 York Avenue, New York, NY
10021-6399}
\vfill
\centerline{\title Abstract}
\bigskip
{\narrower\narrower
We implement the $O(d, d, Z)$ transformations of T-duality
as automorphisms of the operator algebras of Conformal
Field Theories. This extends these transformations to arbitrary
field configurations in the deformation class.\par}
\vfill\vfill\break

T-Duality is a spacetime symmetry of string theory that interchanges
long and short distances
(For a recent review see
\ref\secon{A.\tie Giveon, M.\tie Porrati and E.\tie Rabinovici,
\pre244 (1994).}). It appeared for first time in string theory
in the work of Kikkawa and Yamasaki and of Sakai
and Senda
\ref\first{K.\tie Kikkawa and M.\tie Yamasaki, \pl149
(1984), 357; N.\tie Sakai and I.\tie Senda, Prog. Theor. Phys. {\bf
75} (1986), 692.}. These authors considered a string moving on a
spacetime in which one spatial dimension is compactified on a circle
of radius $R$. In such a background the string possesses two types of
states: momentum (strings with integer quantized
momenta in the compact dimension)
and winding excitations (strings winding around the
compact dimension an integer number of times). The masses of the
momentum excitations are of the form ${}\sim
n/R$, with integral $n$, while those of the winding modes are ${}\sim
m R$, where $m$ is the number of times the string wraps around the
compact dimension. Already, a duality is apparent in which
interchanging the r\^{o}le of the momentum and winding modes is
equivalent to mapping $R\rightarrow 1/R$.

This particular duality is only part of a larger group of discrete symmetries
which arise when we consider strings moving on a $d$-dimensional
toroidal background and in the presence of a constant non-zero
antisymmetric tensor field. These duality transformations
generate  the discrete group $O(d, d, Z)$ and
an additional symmetry
which acts on the background antisymmetric tensor field by
$b_{\mu\nu} \rightarrow -b_{\mu\nu}$ \ref{\eqwvio}{A. Giveon,
E. Rabinovici and G.
Veneziano, \np322 (1989), 167; V. P. Nair, A. Shapere, A. Strominger and
F. Wilczek, \np287 (1987), 402; A. Shapere and
F. Wilczek, \np320 (1989), 669; R. Dijkgraaf,
E. Verlinde and H. Verlinde, in {\it Perspectives
in string theory}, P. Di Vecchia and J. L. Petersen
Eds. (World Scientific, Singapore, 1988); J. Schwarz,
in {\it Elementary Particles and the universe} (1989) 69 and
Caltech prerpint CALT-68-1581; J. Maharana
and J. Schwarz \np390 (1993), 3.}. Dine, Huet and Seiberg
\ref\dhs{M.\tie Dine, P.\tie Huet and N.\tie Seiberg, \np 322 (1989), 301.},
observed some time ago that T-duality is, in fact, a finite gauge
transformation. 
For generic values of the radius of the circle, the unbroken 
gauge symmetry of
the theory is $U(1)_L \times U(1)_R$ due to the Kaluza-Klein mechanism.
It is a surprising result of string theory that this gauge symmetry is
enhanced to $SU(2)_L \times SU(2)_R$ when the radius of the circle
acquires a critical value. T-Duality is then 
a particular $SU(2)_L \times SU(2)_R$ gauge
transformation: more specifically it corresponds
to the global gauge transformation in the
Weyl subgroup of $SU(2)$.
This observation seems more natural when we recall that the
extra gauge bosons are winding modes, while the Kaluza Klein excitations
are momentum modes. The enhanced gauge symmetry therefore mixes
winding and momentum modes, just as T-Duality interchanges them.
Later Giveon, Malkin and Rabinovici \ref\third{
A.\tie Giveon, N.\tie Malkin and E.\tie Rabinovici, \pl238 (1990), 57.}
generalized this result to any $d$-dimensional toroidal background.
They demonstrated that the discrete $O(d, d,Z)$ duality transformations
are particular gauge transformations of the underlying gauge symmetry
of the string background.

In recent years our understanding of gauge symmetries in string theory
has improved considerably \ref{\ovre}{M.\tie Evans and
B.\tie Ovrut, \pr41 (1990), 3149; \pl231 (1989), 80.}, \ref{\eaw}{M. Evans
and I. Giannakis, \pr44 (1991), 2467; M. Evans, I. Giannakis
and D. V. Nanopoulos, \pr50 (1994), 4022.}.
We understand how to implement gauge
transformations on arbitrary backgrounds. Gauge transformations
are generated
by certain similarity transformations of the stress-tensors of the
associated conformal field theories
$$
T(\sigma) \longmapsto e^{ih} T(\sigma) e^{-ih}.
\numbereq\name\eqsim
$$
This yields a new transformed stress-tensor from which we may derive
the transformation properties of the spacetime fields. The transformation
of the stress-tensor is to be expected; stress-tensors are parameterized
by the spacetime field configurations and transforming the
spacetime fields will therefore transform the stress-tensor.

What is then the
operator $h$ that implements gauge transformations? For each
unbroken gauge
symmetry there exists a corresponding current algebra on the
world-sheet. For a gauge transformation
with parameter $\Lambda^a(X)$,
the generator, $h$, in equation (\eqsim) is just
$$
h = \int\; d\sigma \> \Lambda^a(X(\sigma)) J^a(\sigma)
\numbereq\name\eqgen
$$
where $J^a(\sigma)$ are the generators of the current algebra.
We may also generate general coordinate and two-form
gauge transformations by choosing [\eaw]
$$
h={\int}d{\sigma}[{\xi}^{\mu}(X){\partial}X_{\mu}+{\zeta}^{\mu}(X)
{\overline{\partial}}X_{\mu}].  \numbereq\name{\eqabouo}
$$
In reference \ref\for{M.\tie Evans and I.\tie Giannakis hep-th/9511061,
Rockefeller University preprint RU95-04-B.}
this general approach to understanding the symmetries of string theory
was applied to T-Duality. The operator $h$ which implements T-Duality
as an inner automorphism of the operator algebra was identified as
one of the $SU(2)$ currents, as in eq. (\eqgen). In this letter we will
apply our understanding of gauge symmetries in string theory with the insight
of GMR [\third] that $O(d, d, Z)$ transformations are gauge
transformations, and we will identify the operators $h$, Eq. (\eqgen),
which implement these transformations.
These operators map the operator algebra onto itself and in addition
can be pulled back to spacetime and be interpreted as $O(d, d,Z)$
transformations on the spacetime fields. This will be achieved by
fixing the operator algebra and constructing the operators $h$ at one
of the self-dual points. The effect of an $O(d, d, Z)$ transformation
on arbitrary spacetime fields can then
be calculated by applying the same inner
automorphisms to general stress-tensor in the deformation class.

For simplicity we shall initially consider a string moving on a two-dimensional
torus parameterized by a constant metric $g_{\mu\nu}$ and antisymmetric
tensor field $b_{\mu\nu}$. Thus spacetime is $M^{24} \times T^2$.
The stress-tensor of the corresponding Toroidal
Conformal Field Theory is
$$
T_{g,b}(\sigma)={1\over 2}g^{\mu\nu}:{\hat{\partial
X_{\mu}}}{\hat{\partial X_{\nu}}}:(\sigma)
\qquad {\overline T}_{g,b}(\sigma)={1\over 2}g^{\mu\nu}:
\hat{{\overline\partial}
X_{\mu}}\hat{{\overline\partial}X_{\nu}}:(\sigma),
\numbereq\name{\eqsytu}
$$
where
$$
{\hat{\partial X_{\mu}}}(\sigma) ={1\over {\sqrt 2}}({\pi}_{\mu}
+(g_{\mu\nu}+b_{\mu\nu})X'^{\nu})(\sigma), \qquad
\hat{{\overline\partial}X_{\mu}}(\sigma)=
{1\over {\sqrt 2}}({\pi}_{\mu}-(g_{\mu\nu}+b_{\mu\nu})X'^{\nu})(\sigma)
\numbereq\name{\eqsmz}
$$
and $\mu, \nu=1, 2$.
At one of the symmetric points of the deformation class the generic
gauge symmetry
$[U(1)_L]^2 \times [U(1)_R]^2$ is enhanced to $[SU(2)_L]^2
\times [SU(2)_R]^2$. 
At this point the stress-tensor is
$$
T_G(\sigma)={1\over 2}G^{\mu\nu}:{\partial}
X_{\mu}{\partial}X_{\nu}:(\sigma),
\qquad {\overline T}_G(\sigma)={1\over 2}G^{\mu\nu}:{\overline{\partial}}
X_{\mu}{\overline{\partial}}X_{\nu}:(\sigma)
\numbereq\name{\eqsoaz}
$$
where $G_{\mu\nu}$ is a constant diagonal metric (the identity!) and the
antisymmetric background field has been set to zero.
Throughout this paper, we shall use ${\partial}X_{\mu}(\sigma)$
and ${\overline{\partial}}X_{\mu}(\sigma)$ (without the hat) to
denote the light-cone derivatives at the critical point, i.e.
$$
{\partial}X_{\mu}(\sigma)={1\over {\sqrt 2}}({\pi}_{\mu}
+G_{\mu\nu}X'^{\nu})(\sigma), \qquad {\overline{\partial}}X_{\mu}(\sigma)=
{1\over {\sqrt 2}}({\pi}_{\mu}-G_{\mu\nu}X'^{\nu})(\sigma)
\numbereq\name{\eqszpa}
$$
irrespective of the actual backgrounds $g_{\mu\nu}$ and $b_{\mu\nu}$.
The stress-tensor has been defined through a point
splitting regularization as follows
$$
T_{g, b}(\sigma)={\textstyle{1\over 2}}g^{\mu\nu}:\hat{\partial X}_{\mu}
\hat{\partial X}_{\nu}:(\sigma)
={\lim_{{\epsilon}\to 0}}
{\textstyle{1\over 2}}g^{\mu\nu}\hat{\partial X}_{\mu}
(\sigma)\hat{\partial X}_{\nu}(\sigma+\epsilon)
+{{g^{\mu\nu}g_{\mu\nu}}\over {4{\pi}{\epsilon}^2}}.
\numbereq\name{\eqdoye}
$$
We should now elaborate on a technical point that can nonetheless
be very confusing. The operator ${\hat{\partial X_{\mu}}}(\sigma)$
in equation (\eqsytu) is not the same operator for different
string backgrounds.  As we deform our CFT by varying the spacetime fields,
${\hat{\partial X_{\mu}}}(\sigma)$ changes. This is
also apparent from equations
(\eqsmz) and (\eqszpa). Because we want to
compare CFT's at different points
of the deformation class we need to express our operators in terms of a fixed
basis. We will choose to fix our operator algebra at one of the
critical points of the deformation class (the $SU(2)$ point), meaning that
we will fix the commutation relations of
$\pi_{\mu}(\sigma)$ and $X^{\nu}(\sigma)$ to be
$$
[\pi_{\mu}(\sigma), X^{\nu}(\sigma')]=i{\delta}_{\mu}^{\nu}
{\delta}({\sigma}-{\sigma'}).
\numbereq\name{\eqfasoul}
$$
The operators $\pi_{\mu}(\sigma)$ and $X^{\nu}(\sigma)$ obey
fixed commutation relations everywhere in
the deformation class, independent
of the spacetime fields while ${\hat{\partial X_{\mu}}}(\sigma)$
do not. Having done that we can express the stress tensor of a
generic point in terms of these operators. It turns out to be more
convenient to express ${\hat{\partial X_{\mu}}}(\sigma)$ in terms
of ${\partial}X_{\mu}(\sigma)$ 
and  ${\overline{\partial}}X_{\mu}(\sigma)$ as follows
$$
{\hat{\partial X_{\mu}}}={1\over 2} {\lbrack ({\partial X_{\mu}}+
{\overline{\partial}}X_{\mu})+(g_{\mu\rho}+b_{\mu\rho})G^{\rho\nu}
({\partial X_{\nu}}-{\overline{\partial}}X_{\nu}) \rbrack}.
\numbereq\name{\eqauios}
$$
Substituting into Eq. (\eqsytu) we get
$$
\eqalign{
&T_{g,b}(\sigma)={1\over 8} \lbrack
(g^{\mu\nu}+g^{\mu\rho}(g_{\rho\sigma}
+b_{\rho\sigma})G^{\sigma\nu}+g^{\rho\nu}(g_{\rho\sigma}
+b_{\rho\sigma})G^{\mu\sigma} +g^{\rho\sigma}(g_{\rho\kappa}
+b_{\rho\kappa})(g_{\sigma\lambda}+b_{\sigma\lambda})\cr
&G^{\mu\kappa}G^{\lambda\nu}){\partial X_{\mu}}(\sigma)
{\partial X_{\nu}}(\sigma+\epsilon)
+(g^{\mu\nu}-g^{\mu\rho}(g_{\rho\sigma}
+b_{\rho\sigma})G^{\sigma\nu}-g^{\rho\nu}(g_{\rho\sigma}
+b_{\rho\sigma})G^{\mu\sigma}+g^{\rho\sigma}(g_{\rho\kappa}\cr
&+b_{\rho\kappa})
(g_{\sigma\lambda}+b_{\sigma\lambda})
G^{\mu\kappa}G^{\lambda\nu}){\overline{\partial}}X_{\mu}(\sigma)
{\overline{\partial}}X_{\nu}(\sigma+\epsilon)
+(g^{\mu\nu}-g^{\rho\sigma}(g_{\rho\kappa}
+b_{\rho\kappa})(g_{\sigma\lambda}+b_{\sigma\lambda})
G^{\mu\kappa}\cr
&G^{\lambda\nu})
({\partial X_{\mu}}(\sigma){\overline{\partial}}
X_{\nu}(\sigma+\epsilon)+{\partial X_{\mu}}(\sigma+\epsilon)
{\overline{\partial}}X_{\nu}(\sigma))+{{g^{\mu\nu}
g_{\mu\nu}}\over {4{\pi}{\epsilon}^2}}
 \rbrack \cr}
\numbereq\name{\eqcerou}
$$
and a similar expression for ${\overline T}_{g, b}$. We have
expressed the stress-tensor of a generic point of the
deformation class in terms of a fixed basis of operators at
the self-dual point.

The discrete symmetry group of the $d=2$ compactifications consists
of $O(2, 2, Z)$ transformations and $b_{12} \rightarrow
-b_{12}$. The $O(2, 2, Z)$ group is generated by
permutations, reflections
and some particular linear transformations of the coordinates as well as
factorized dualities
($R \rightarrow {1\over R}$ type of transformations) and
integer shifts of the antisymmetric tensor [\third].
Factorized dualities interchange the role of   $\pi_{\mu}(\sigma)$
and $X'^{\nu}(\sigma)$ and so we seek operators $h^{i}$ that
can achieve this; we need
$$
e^{ih^{i}}{\pi}_{\mu}(\sigma)e^{-ih^{i}}=
-G_{\mu\nu}X'^{\nu}(\sigma) \qquad e^{ih^{i}}
G_{\mu\nu}X'^{\nu}(\sigma)
e^{-ih^{i}}=-{\pi}_{\mu}(\sigma),
\numbereq\name{\eqcoua}
$$
 which from the definition (\eqszpa) is equivalent to,
$$
e^{ih^{i}}{\partial X}_{\mu}(\sigma)
e^{-ih^{i}}=-{\partial X}_{\mu}(\sigma) \qquad e^{ih^{i}}
{\overline{\partial}}X_{\mu}(\sigma)e^{-ih^{i}}
={\overline{\partial}}X_{\mu}(\sigma).
\numbereq\name{\eqcima}
$$
To find these operators $h^{i}$, we recal that at the symmetric point of
the {\it deformation class} the gauge symmetry of the theory is
 $[SU(2)_L]^2 \times [SU(2)_R]^2$. This symmetry enhancement
is due to the appearance of extra $(1,0)$ and
$(0,1)$ operators $e^{{\pm i}{\sqrt 2}X_L^{1, 2}}(\sigma)$, $e^{{\pm i}
{\sqrt 2}X_R^{1, 2}}(\sigma)$. The operators $\partial X_{1, 2}(\sigma),
e^{{\pm i}{\sqrt 2}X^{1, 2}_L}$ then form an $[SU(2)_L]^2$ algebra.
It is then straightforward to
construct the operators $h^{i}$ as follows
$$
h^{(i)}={1\over 2i}{\int}d{\sigma}{\Lambda}_{i}
(e^{i{\sqrt 2}k^{(i)}_\mu X^{\mu}_L}-
e^{-i{\sqrt 2}k^{(i)}_\mu X^{\mu}_L}),
\numbereq\name{\eqaoyp}
$$
where $k^{(i)}_\mu$ is a suitable basis of Killing forms on the
torus. For a $D$-dimensional flat torus, there are $D$ of them $(i
=1, \cdots, D)$ and we have chosen a particular basis where $k^{(i)}=
(1,0,0, \cdots), (0,1,0, \cdots), (0,0,1, \cdots), \ldots $.

The effect of these inner automorphisms on  
${\partial X_{\mu}}$ and ${\overline{\partial}}X_{\mu}$ at the critical
point  can be calculated by writing
$$
e^{ih^{i}}{\partial X}_{\mu}(\sigma)
e^{-ih^{i}}
={\partial X}_{\mu}(\sigma)
+i{\lbrack h^{i}, {\partial X}_{\mu}(\sigma) \rbrack}
+{1\over 2}{\lbrack h^{i}, {\lbrack h^{i}, {\partial X}_{\mu}
(\sigma) \rbrack},
\rbrack}+\cdots
\numbereq\name{\eqasopu}
$$
and a similar expression for ${\overline{\partial}}X_{\mu}$ and
calculating the commutators explicitly. We find that
$$
e^{ih^{1, 2}}{\partial X}_{1, 2}(\sigma)
e^{-ih^{1, 2}}={\partial X}_{1, 2}(\sigma){\cos{\Lambda_{1, 2}}}-
{1\over {\sqrt 2}}(e^{i{\sqrt 2}{X_L}^{1, 2}}+
e^{-i{\sqrt 2}{X_L}^{1, 2}}){\sin{\Lambda_{1, 2}}}
\numbereq\name{\eqxosz}
$$
and
$$
e^{ih^{1, 2}}{\overline{\partial}}X_{1, 2}(\sigma)
e^{-ih^{1, 2}}={\overline{\partial}}X_{1, 2}(\sigma).
\numbereq\name{\eqcsca}
$$
We observe that if we choose
$\Lambda_1=\Lambda_2=\pi$ this particular
automorphism satisfies equation (\eqcima).
This result depends solely on the $[SU(2)_L]^2 \times [SU(2)_R]^2$
algebra. If our compact space is a $d$-dimensional torus, then there
are $d$ separate T-dualities. We can consider the effect of the
product of these dualities on the
stress-tensor Eq. (\eqcerou). With some
amount of algebra we can see that the transformed
stress-tensor is of the same form with the original, but with
transformed {\it space-time} fields
$$
e^{ih^{(1)}}e^{ih^{(2)}}
T_{g,b}e^{-ih^{(2)}}e^{-ih^{(1)}}(\sigma)=T_{{\tilde g},
{\tilde b}} ,
\numbereq\name{\eqcerq}
$$
where
$$
{\tilde g}^{\mu\nu}+{\tilde b}^{\mu\nu}=(g_{\kappa\lambda}
+b_{\kappa\lambda})G^{\kappa\mu}G^{\lambda\nu}.
\numbereq\name{\eqasopi}.
$$
We can also consider the effect of a separate T-duality generated
by $h^{1}$ for example.  The effect of this separate automorphism is
to change the sign of $\partial X_1 (\sigma)$ and leave
$\partial X_2(\sigma)$ invariant. Acting on the stress-tensor again 
with this separate automorphism will
produce again a new stress-tensor
$$
e^{ih^{(1)}}
T_{g,b}e^{-ih^{(1)}}(\sigma)=T_{{\tilde g},
{\tilde b}} ,
\numbereq\name{\eqcjherq}
$$
where the space-time fields transform as follows
$$
\eqalign{
{\tilde g}^{11}&=g_{11}G^{11}G^{11}
+g^{22}b_{21}b_{12}G^{11}G^{11} \cr
{\tilde g}^{12}&=g^{22}b_{12}
G^{11} \cr} \qquad \eqalign{{\tilde g}^{22}&=g^{22}
\cr
{\tilde b}_{12}&=G_{11}{{g^{12}}\over {g^{22}}}. \cr}
\numbereq\name{\eqxesis}
$$
These transformations were first derived by Buscher, \ref\buscher
{T. Buscher, \pl 159 (1985), 127,
\pl 194 (1987), 59 and \pl 201 (1988), 466.}. One comment is in
order here. The choice of the operators $h^i$ which implement T-duality,
eq. (\eqaoyp), is not unique. Instead, $h^i$ could
have been chosen as
$$
h^{(i)}={1\over 2}{\int}d{\sigma}{\Lambda_i}
(e^{i{\sqrt 2}k^{(i)}_\mu X^{\mu}_L}+
e^{-i{\sqrt 2}k^{(i)}_\mu X^{\mu}_L}).
\numbereq\name{\eqaod}
$$
The effect of these automorphisms on ${\partial X_{\mu}}$
and ${\overline{\partial}}X_{\mu}$
can be calculated to be
$$
e^{ih^{1, 2}}{\partial X}_{1, 2}(\sigma)
e^{-ih^{1, 2}}={\partial X}_{1, 2}(\sigma){\cos{\Lambda_{1, 2}}}-
{1\over {\sqrt 2}}(e^{i{\sqrt 2}{X_L}^{1, 2}}-
e^{-i{\sqrt 2}{X_L}^{1, 2}}){\sin{\Lambda_{1, 2}}}
\numbereq\name{\eqxosz}
$$
and
$$
e^{ih^{1, 2}}{\overline{\partial}}X_{1, 2}(\sigma)
e^{-ih^{1, 2}}={\overline{\partial}}X_{1, 2}(\sigma)
\numbereq\name{\eqcsca}
$$
and it leads to the same transformation properties for
$g_{\mu\nu}$ and $b_{\mu\nu}$ for $\Lambda_1=\Lambda_2=\pi$.
The choice of the appropriate operators acquires significance
in the $Z_2$ orbifold models in which only one of the two survives the
$Z_2$ projection. 

The constant shifts of the antisymmetric tensor $b_{12} \rightarrow
b_{12}+{\theta}_{12}$ correspond to a particular two-form
gauge transformations. Two-form gauge invariance in string
theory is responsible for the extension of the $U(1)$ symmetry
which appears when one compactifies any generally covariant
theory on a circle to $U(1)_L \times U(1)_R$. The operator $h$
which implements generic two-form gauge transformations
is given by
$$
h=\ints\zeta^\mu(X)\left(\dx_\mu-\dbx_\mu\right).
\numbereq\name{\eqhex}
$$
Let's then choose our parameter of transformation to be
$\zeta^\mu(X)={\theta^{\mu}_{\nu}}X^\nu$ where ${\theta^{\mu\nu}}=-
{\theta^{\nu\mu}}$.
The action of the inner automorphism generated by this particular choice
of the parameter on ${\partial X_{\mu}}$
and ${\overline{\partial}}X_{\mu}$
can be calculated to be
$$
\eqalign{
e^{ih}{\partial X}_{\mu}(\sigma)
e^{-ih}&={\partial X}_{\mu}(\sigma)+{\theta_{\mu\nu}}G^{\nu\rho}
\left(\dx_\rho-\dbx_\rho\right)\cr
e^{ih}
{\overline{\partial}}X_{\mu}(\sigma)
e^{-ih}&={\overline{\partial}}X_{\mu}(\sigma)+{\theta_{\mu\nu}}G^{\nu\rho}
\left(\dx_\rho-\dbx_\rho\right). \cr}
\numbereq\name{\eqcimura}
$$
The operator $e^{ih}$ needs to be single-valued when
$X^{\mu} \mapsto X^{\mu}+2{\pi}R{\eta}^{\mu}$. This
implies then that the
parameters $\theta_{\mu\nu}$ are integers. 
Subsequently the effect of this automorphism on the the stress-tensor
Eq. (\eqcerou) reads
$$
\eqalign{
&e^{ih}T_{g,b}(\sigma)e^{-ih}={1\over 8}  \lbrack
(g^{\mu\nu}+g^{\mu\rho}(g_{\rho\sigma}
+b_{\rho\sigma}+{\theta_{\rho\sigma}})G^{\sigma\nu}
+g^{\rho\nu}(g_{\rho\sigma}
+b_{\rho\sigma}+{\theta_{\rho\sigma}})
G^{\mu\sigma} \cr
&+g^{\rho\sigma}(g_{\rho\kappa}
+b_{\rho\kappa}+{\theta_{\rho\kappa}})
(g_{\sigma\lambda}+b_{\sigma\lambda}+{\theta_{\sigma\lambda}})
G^{\mu\kappa}G^{\lambda\nu}){\partial X_{\mu}}(\sigma)
{\partial X_{\nu}}(\sigma+\epsilon)
+(g^{\mu\nu}-g^{\mu\rho}(g_{\rho\sigma}\cr
&+b_{\rho\sigma}+{\theta_{\rho\sigma}})
G^{\sigma\nu}
-g^{\rho\nu}(g_{\rho\sigma}
+b_{\rho\sigma}+{\theta_{\rho\sigma}})
G^{\mu\sigma}+g^{\rho\sigma}(g_{\rho\kappa}
+b_{\rho\kappa}+{\theta_{\rho\kappa}})
(g_{\sigma\lambda}+b_{\sigma\lambda}+{\theta_{\sigma\lambda}})\cr
&G^{\mu\kappa}
G^{\lambda\nu}){\overline{\partial}}X_{\mu}(\sigma)
{\overline{\partial}}X_{\nu}(\sigma+\epsilon)
+(g^{\mu\nu}-g^{\rho\sigma}(g_{\rho\kappa}
+b_{\rho\kappa}+{\theta_{\rho\kappa}})(g_{\sigma\lambda}+
b_{\sigma\lambda}+{\theta_{\sigma\lambda}})
G^{\mu\kappa}G^{\lambda\nu})\cr
&({\partial X_{\mu}}(\sigma)
{\overline{\partial}}X_{\nu}(\sigma+\epsilon)
+{\partial X_{\mu}}(\sigma+\epsilon){\overline{\partial}}X_{\nu}(\sigma)
)+{{g^{\mu\nu}
g_{\mu\nu}}\over {4{\pi}{\epsilon}^2}} \rbrack=T_{g, b+{\theta}}. \cr}
\numbereq\name{\eqpinou}
$$
Thus the inner automorphism generated by $h$ maps the world-sheet
stress-tensor onto a different one. The resulting CFT is isomorphic to
the original one and this particular automorphism can be interpreted
as a transformation on the spacetime fields 
$$
g_{\mu\nu} \rightarrow {\tilde g_{\mu\nu}}=g_{\mu\nu}, \qquad
b_{\mu\nu} \rightarrow {\tilde b_{\mu\nu}}=b_{\mu\nu}+{\theta_{\mu\nu}}.
\numbereq\name{\eqxouzps}
$$
Similarly permutations of the coordinates are
particular coordinate transformations
(rotations). A generic coordinate transformation is generated by
$$
h=\ints\xi^\mu(X)\left(\dx_\mu+\dbx_\mu\right).
\numbereq\name{\eqhzox}
$$
If we choose now $\xi^\mu(X)=\omega^{\mu}_{\nu}X^\nu$
($\omega^{\mu\nu}=-\omega^{\nu\mu}={\theta}
{\epsilon^{\mu\nu}}$) we find that
$$
\eqalign{
e^{i{h}}{\partial X}_{\mu}(\sigma)
e^{-ih}&={\cos{\theta}}{\partial X}_{\mu}(\sigma)+
{\sin{\theta}}{\epsilon_{\mu\nu}}G^{\nu\rho}\dx_\rho(\sigma) \cr
e^{ih}
{\overline{\partial}}X_{\mu}(\sigma)
e^{-ih}&={\cos{\theta}}{\overline{\partial}}X_{\mu}(\sigma)
+{\sin{\theta}}{\epsilon_{\mu\nu}}G^{\nu\rho}\dbx_\rho(\sigma). \cr}
\numbereq\name{\eqcimurou}
$$
Permutations of the coordinates correspond to $\theta={{\pi}\over 2}$.
Then this automorphism upon acting on the stress-tensor
Eq. (\eqcerou) will produce a new transformed stress-tensor where the
transformed spacetime fields will read
$$
\eqalign{
{\tilde g}_{11}&=g_{22}{\epsilon_{21}}{\epsilon_{21}}G^{22}
G^{22} \cr
{\tilde g}_{12}&=g_{12}{\epsilon_{12}}{\epsilon_{21}}G^{11}
G^{22} \cr} \qquad \eqalign{{\tilde g}_{22}&=g_{11}
{\epsilon_{21}}{\epsilon_{21}}G^{11}
G^{11} \cr
{\tilde b}_{12}&=b_{12}. \cr} \numbereq\name{\eqsxeseis}
$$
Reflections of the coordinates correspond to operators $h$ which
act on ${\partial X_{\mu}}$ and ${\overline{\partial}}X_{\mu}$ as follows
 $$
e^{ih}{\partial X}_{\mu}(\sigma)
e^{-ih}=-{\partial X}_{\mu}(\sigma) \qquad e^{ih}
{\overline{\partial}}X_{\mu}(\sigma)
e^{-ih}=-{\overline{\partial}}X_{\mu}(\sigma).
\numbereq\name{\eqcimaks}
$$
This can be achieved if we choose $h=h_1+h_2$
with 
$$
h_1^{i}={1\over 2i}{\int}d{\sigma}{\Lambda_i}
(e^{i{\sqrt 2}k^{(i)}_\mu X^{\mu}_L}-
e^{-i{\sqrt 2}k^{(i)}_\mu X^{\mu}_L})
\qquad h_2^{i}={1\over 2i}{\int}
d{\sigma}{\tilde{\Lambda}}_i(e^{i{\sqrt 2}k^{(i)}_\mu X^{\mu}_R}-
e^{-i{\sqrt 2}k^{(i)}_\mu X^{\mu}_R})
\numbereq\name{\eqalx}
$$
and $\Lambda_i={\tilde\Lambda}_i=\pi$. Acting again with
this automorphism on the
stress tensor Eq. (\eqcerou) we obtain a modified
stress-tensor $T_{{\tilde g},
{\tilde b}}$ with
$$
{\tilde g}^{11}=g^{11}, \qquad {\tilde g}^{22}
=g^{22}, \qquad {\tilde g}^{12}=
-g^{12}, \qquad {\tilde b}_{12}=-b_{12}.
\numbereq\name{\eqazobow}
$$
Finally the remaining $O(d, d, Z)$ transformations
correspond to coordinate
transformations and subsequently are generated
by $h$, Eq. (\eqhzox) with
$\xi^{\mu}=(0, {\epsilon}^{2}_{1}X^1)$.
The effect of this automorphism on  ${\partial X_{\mu}}$
and ${\overline{\partial}}X_{\mu}$
can be calculated as follows
$$
\eqalign{
e^{ih}{\partial X}_{1}(\sigma)
e^{-ih}&={\partial X}_{1}+{{\epsilon_1^2}\over 2}({\partial X}_{2}+
{\overline{\partial}}X_{2}) \cr
e^{ih}{\overline{\partial}}X_{1}(\sigma)
e^{-ih}&={\overline{\partial}}X_{1}
+{{\epsilon_1^2}\over 2}({\partial X}_{2}+
{\overline{\partial}}X_{2}) \cr
e^{ih}{\partial X}_{2}(\sigma)
e^{-ih}&={\partial X}_{2}+
G_{22}G^{11}{{\epsilon_1^2}\over 2}({\partial X}_{1}-
{\overline{\partial}}X_{1}) \cr
e^{ih}{\overline{\partial}}X_{2}(\sigma)
e^{-ih}&={\overline{\partial}}X_{2}
-G_{22}G^{11}{{\epsilon_1^2}\over 2}({\partial X}_{1}-
{\overline{\partial}}X_{1}). \cr}
\numbereq\name{\eqasohd}
$$
Having these relations at our disposal we proceed to calculate the
effect of this automorphism on the stress tensor which describes string
propagation on $T^2$ at a generic point of the deformation class.
The resulting stress tensor takes an intimidating form but with some tedius
algebra we realize that the new stress tensor corresponds to the original
one with the following transformed {\it space-time fields}
$$
{\tilde g}^{11}=g^{11}, \qquad {\tilde g}^{22}=g^{22}+
{\epsilon_1^2}{\epsilon_1^2}g^{11}-2{\epsilon_1^2}g^{12}
, \qquad {\tilde g}^{12}=
g^{12}-{\epsilon_1^2}g^{11}, \qquad {\tilde b}_{12}=b_{12}.
\numbereq\name{\eqazobojsw}
$$

The remaining discrete symmetry of string theory compactified on
a $d$-dimensional torus corresponds to the transformation
$b_{\mu\nu} \rightarrow -b_{\mu\nu}$.
In order to implement it as an automorphism of the operator algebra
we seek an operator $h$ whose action on the stress-tensor can
be interpreted as this particular spacetime transformation
$$
e^{ih}T_{g,b}e^{-ih}(\sigma)=T_{{\tilde g},
{\tilde b}} =T_{g, -b},
\numbereq\name{\eqfdcerq}
$$
Then the operator $h$, in order to be interpreted
as implementing this specific
symmetry transformation, needs to act on  
${\partial X_{\mu}}$ and ${\overline{\partial}}X_{\mu}$ in the following
manner
$$
e^{ih}{\partial X}_{\mu}(\sigma)
e^{-ih}={\overline{\partial}}X_{\mu}(\sigma), \qquad
e^{ih}{\overline{\partial}}X_{\mu}(\sigma)
e^{-ih}={\partial X}_{\mu}(\sigma).
\numbereq\name{\eqleri}
$$
So we seek an automorphism that interchanges the two Virasoro algebras.
Clearly such an automorphism will correspond to an outer
automorphism since no inner automorphism can
achieve this: the algebra is
a tensor product of a left and a right Virasoro.

In this paragraph we shall summarize what we
have done in this paper. We
have applied a general approach to understanding
gauge symmetries in string theories as automorphisms of
the operator algebra to $O(d, d, Z)$ dualities. We exhibited explicitly
all the operators that implement the inner automorphisms of
the algebra and by applying them to the stress-tensor of the
theory we were able to derive the $O(d, d,Z)$ transformations on
the spacetime fields. The requirement that the operators
that implement the automorphisms are single-valued constrains
the parameters of transformations to be integers.
Although we have worked with a two-dimensional
toroidal background $T^2$ generalization to $d$-dimensions
is straightforward. We have also demonstrated that factorized
dualities and reflections are enhanced gauge symmetries
and as such stringy in nature
while the remaining $O(d, d, Z)$ transformations
are abelian gauge symmetries.

I would like to thank M. Evans and J. Liu for useful discussions.
This work was supported in part by the Department of Energy Contract
Number DE-FG02-91ER40651-TASKB. 

\immediate\closeout1
\bigbreak\bigskip

\line{\twelvebf References. \hfil}
\nobreak\medskip\vskip\parskip

\input refs

\vfil\end